\documentclass[9pt,twocolumn,twoside]{arxivversion}

\journal{ao} 
\setboolean{shortarticle}{false}

\usepackage[mode=buildnew]{standalone}
\usepackage{tikz}
\usepackage{fp,tikz,tikz-3dplot}
\usepackage{tkz-euclide}
\usetikzlibrary{shapes,arrows}   
\usepackage{amsfonts} 
\usetikzlibrary{decorations.pathreplacing,decorations.markings}
\usetikzlibrary{quotes,angles}
\usetikzlibrary{patterns,3d,calc}
\usepackage{comment}

\newcommand{\bvec}[1]{\boldsymbol{\mathbf{#1}}}

\usepackage[utf8]{inputenc} 
\usepackage[T1]{fontenc}
\usepackage{graphicx}
\usepackage{grffile}
\usepackage{mathtools}   
\usepackage{longtable}
\usepackage{wrapfig}
\usepackage{rotating}
\usepackage{grffile} 
\usepackage[normalem]{ulem}
\usepackage{amsmath}
\usepackage{textcomp}

\usepackage{cuted}

\usepackage{capt-of}
\usepackage{hyperref}
\usepackage{subcaption}
\usepackage{marginnote}
\usepackage{setspace}

\usepackage{float}
\usepackage{cases}

\usepackage{physics}
\usepackage{amsmath}
\usepackage{mathtools}

\usepackage{epstopdf}

 \title{Ray-transfer functions for camera simulation of 3D scenes with hidden lens design}

\author[1,*]{Thomas Goossens}
\author[1]{Zheng Lyu}
\author[2]{Jamyuen Ko}
\author[2]{Gordon Wan}
\author[1]{Joyce Farrell}
\author[1]{Brian Wandell}

\affil[1]{Stanford Center for Image Systems Engineering, Stanford University, California 94305, USA}
\affil[2]{Google, Mountain View, California, 94043, USA}
\affil[*]{Email: contact@thomasgoossens.be}

\begin{abstract}
Combining image sensor simulation tools (e.g., ISETCam) with physically based ray tracing (e.g., PBRT) offers possibilities for designing and evaluating novel imaging systems as well as for synthesizing physically accurate, labeled images for machine learning. One practical limitation has been simulating the optics precisely: Lens manufacturers generally prefer to keep lens design confidential. We present a pragmatic solution to this problem using a black box lens model in Zemax; such models provide necessary optical information while preserving the lens designer's intellectual property. First, we describe and provide software to construct a polynomial ray transfer function that characterizes how rays entering the lens at any position and angle subsequently exit the lens. We implement the ray-transfer calculation as a camera model in PBRT and confirm that the PBRT ray-transfer calculations match the Zemax lens calculations for edge spread functions and relative illumination. 
\end{abstract}
\setboolean{displaycopyright}{false}
\begin{document}

\maketitle

\section{Introduction}
In many engineering disciplines, accurate simulation tools enable low-cost and rapid design (soft prototyping). To effectively support image systems simulation, tools must incorporate the full pipeline: the scene, camera optics, and image sensor. In addition to soft prototyping, image systems simulation is increasingly valuable for training machine learning algorithms. In the case of autonomous driving, simulation is being used to create training sets for challenging conditions \cite{omniversevalidation-Kamel2021,liu2020neural, liu2021isetauto,anyverse-mainpage, waymo-simulationcity}. 

There has been extensive work in computer graphics on three-dimensional scene simulation \cite{Goral1984-wd,Sumanta-qr}. There is also a substantial history in simulating optics and sensor simulation \cite{Farrell2003-rb, Farrell2012}. This article addresses an important challenge that has limited the ability to accurately simulate the optics as part of the prototyping.

There is a practical limitation in simulating certain devices: lens manufacturers rarely disclose details of the lens design.  Knowledge of the lens properties is necessary for most lens calculations, and it is essential for accurate soft prototyping and machine learning applications.  Lens manufacturers are sometimes willing, however, to provide a 'black box' description of the optical system. Such black box data can be used with lens design software, such as Zemax OpticStudio, to analyze lens properties. 

A common approach for simulating unknown camera optics makes use of point spread functions (PSF) at different field heights, object distances, and wavelengths \cite{Maeda2005-uf,Short1999,Chen2009a,Farrell2012,Toadere2013,Kim2021a}.
This analysis takes into account geometrical, wave, and chromatic aberrations. An important development was learning how to interpolate the point spread correctly from a finite number of sampled PSFs.  This was demonstrated by \cite{Lehmann2019} and used by \cite{Tseng2021-kt}. 

But important phenomena, such as ray occlusions (which change the PSF) or multiple inter-reflections, are not simulated from the PSF alone; at best, the PSF can be calculated for multiple depths. Occlusions and inter-reflections are present and visually significant in many scenes. Camera simulation of such complex three-dimensional spectral scenes is now practical due to advances in physically-based rendering and increased computing power \cite{pharr2016physically,nimier2019mitsuba,Lyu2021a}. Combining image sensor simulation tools (e.g., ISETCam \cite{isetcam}) with physically based rendering offers new possibilities for designing and evaluating imaging systems. Simulation can also be used to generate large numbers of training scenes along with accurate pixel-levels labels. The ability to synthesize labeled and physically meaningful rendered scenes can significantly accelerate the training cycle for machine learning vision applications \cite{blasinski2018optimizing,liu2019soft,liu2021isetauto}.

In this paper we describe how to use Zemax black box data to obtain an equivalent lens model - the ray transfer function - 
using polynomial approximations \cite{hullin2012polynomial, schrade2016, Zheng2017}.
The ray transfer function, when embedded into a ray-tracer, accounts for three-dimensional effects (depth, occlusion). We evaluate the accuracy of the ray-transfer approximation using two image metrics: relative illumination and blurring. Finally, we provide a link to an open-source and free implementation of all these methods and related data.

\section{Ray-Transfer Matrices and Functions}

\begin{figure}[ht!]
  \begin{subfigure}[t]{0.49\linewidth}
        \captionsetup{justification=centering}
    \caption{\label{fig:rtm}\textbf{Ray-Transfer Matrix}}
  \includegraphics[width=0.99\linewidth]{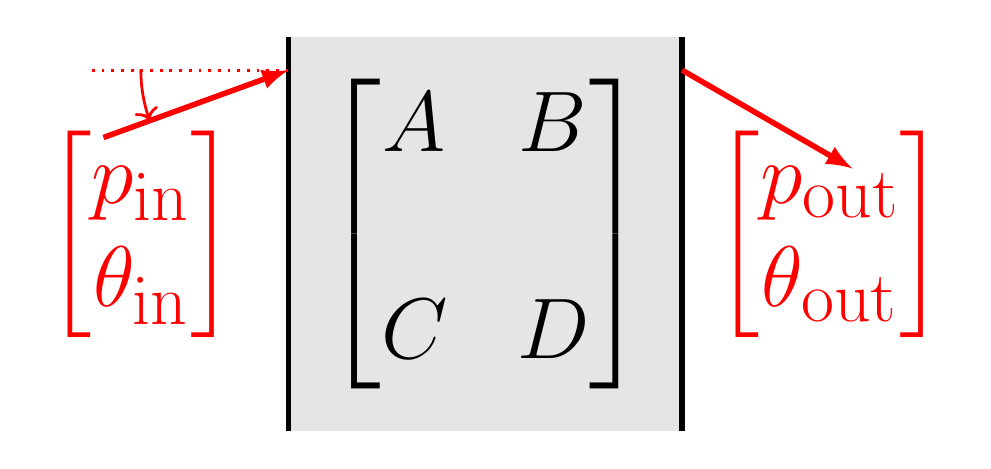}
\end{subfigure}
\begin{subfigure}[t]{0.49\linewidth}
      \captionsetup{justification=centering}
      \caption{\textbf{Ray-Transfer Function (RTF)}}
\includegraphics[width=0.99\linewidth]{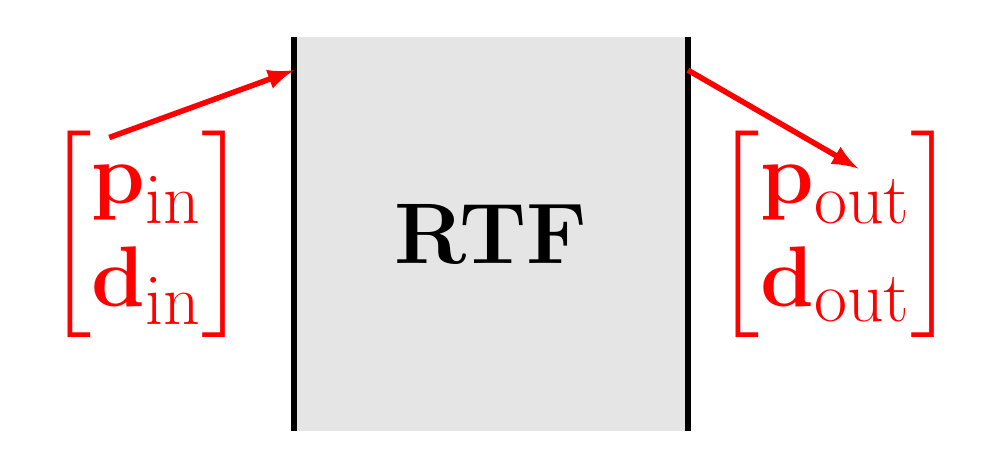}
\end{subfigure}
\centering
\begin{subfigure}[t]{0.49\linewidth}
  \vspace{0.5cm}
  \captionsetup{justification=centering}
    \captionsetup{width=1.1\linewidth}
\caption{\label{fig:rtfcurve}\textbf{RTF with curved output surface}}
\includegraphics[width=0.99\linewidth]{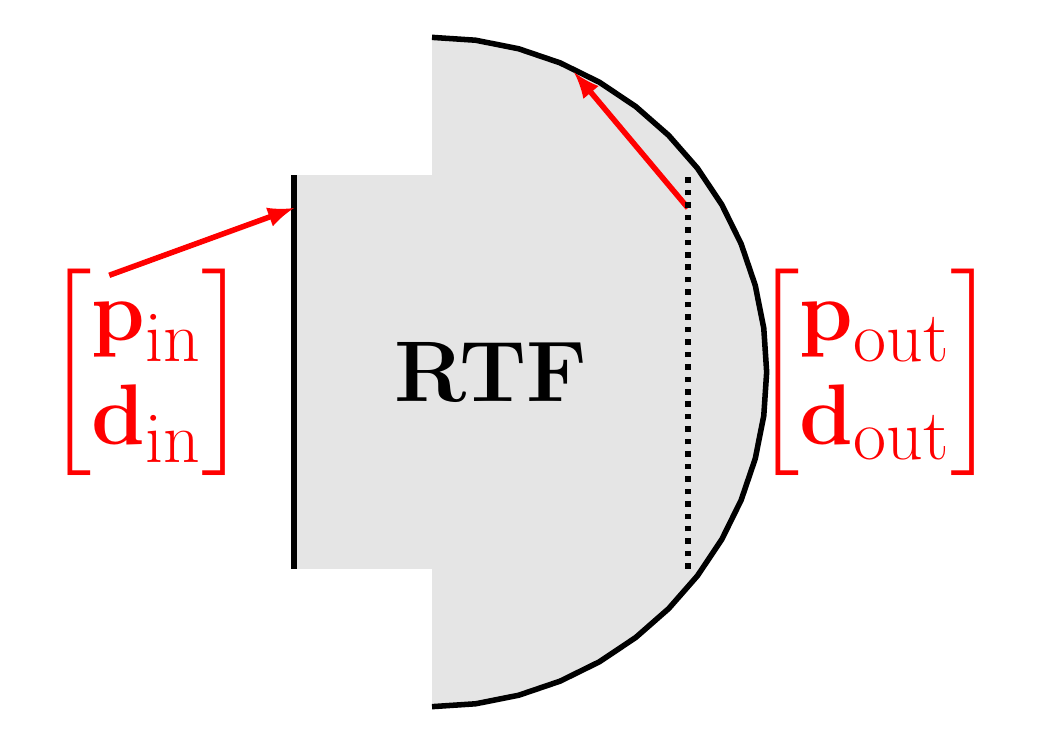}
\end{subfigure}
\caption{\label{fig:rtf} Ray-transfer functions (RTF) extend the classic ABCD ray-transfer matrices used for paraxial calculations. (a) The ray-transfer matrix has position and angle inputs with respect to two planes. (b) The RTF maps a position vector and a unit direction vector (6D) to an output position and direction vector (6D).  (c) The RTF can represent the transformation using curved surfaces, which can be important for wide angle lenses (see Fig.~\ref{fig:wideangle}).}
\end{figure}

For ray-tracing purposes, a lens can be considered a black box system that transforms an incoming ray on one side of the lens to an outgoing ray on another side of the lens. For many lenses, the paraxial ray trace can be accurately modeled using a \emph{Ray-Transfer Matrix} \cite{Goodman2005}.
In this framework, rays are represented by their intersection $p$ at some chosen input and output plane and their angle $\theta$ with respect to the optical axis (see Fig.~\ref{fig:rtm}).
The ray-transfer matrix then relates the angles and intersections at the input and output planes using the well-known ABCD formulation \cite{ABCD-raytransfer-rq} 
\begin{equation}
  \begin{bmatrix}
    p_{\text{out}} \\\theta_{\text{out}}
  \end{bmatrix}
  =
  \begin{bmatrix}
    A & B  \\
    C & D 
  \end{bmatrix}
    \begin{bmatrix}
    p_{\text{in}} \\\theta_{\text{in}}
  \end{bmatrix}.
\end{equation}

In this article, we generalize the matrix to a nonlinear formalism, i.e., \emph{Ray-Transfer Functions} (RTF). The RTF maps incoming rays to outgoing rays in 3D space (Fig.~\ref{fig:rtf}).
A ray is described by an origin vector (in boldface) $\bvec{p}\in\mathbb{R}^3$ and a direction vector $\bvec{d}\in\mathbb{R}^3$ (instead of angles). The transfer function that relates incoming and outgoing rays can be formulated as
\begin{equation}
\label{eq:tf}
  \left[
  \begin{array}{l}
    \\ \bvec{p}_{\text{out}} \\\\
    \hline
    \\ \bvec{d}_{\text{out}} \\\\
  \end{array}\right]
  =   \left[ 
    \begin{array}{c}
     x\\y\\z\\
    \hline
     d_x\\d_y\\d_z
  \end{array}\right]_{\text{out}}
=  \bvec{f}(\bvec{p}_{\text{in}};\bvec{d}_{\text{in}})
=  \left[
  \begin{array}{l}
    f_{x}(\bvec{p}_{\text{in}};\bvec{d}_{\text{in}})\\
    f_{y}(\bvec{p}_{\text{in}};\bvec{d}_{\text{in}})\\
    f_{z}(\bvec{p}_{\text{in}};\bvec{d}_{\text{in}})\\
    \hline
    f_{d_x}(\bvec{p}_{\text{in}};\bvec{d}_{\text{in}})\\
    f_{d_y}(\bvec{p}_{\text{in}};\bvec{d}_{\text{in}})\\
    f_{d_z}(\bvec{p}_{\text{in}};\bvec{d}_{\text{in}})\\
 \end{array}\right].
\end{equation}

\begin{figure}[ht!]
\begin{subfigure}[t]{0.49\linewidth}
  \centering
  \caption{\label{fig:dgauss28deg} Double Gauss 28deg.}
  \includegraphics[width=0.99\linewidth]{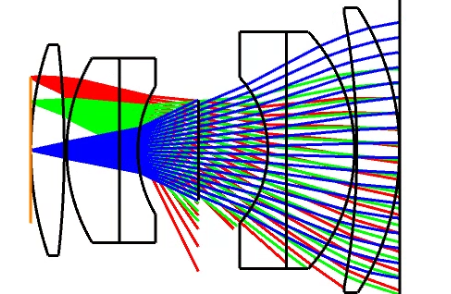}

\end{subfigure}
  \begin{subfigure}[t]{0.49\linewidth}
  \caption{\label{fig:wideangle} Wide angle lens 200deg. }
    \includegraphics[width=0.99\linewidth]{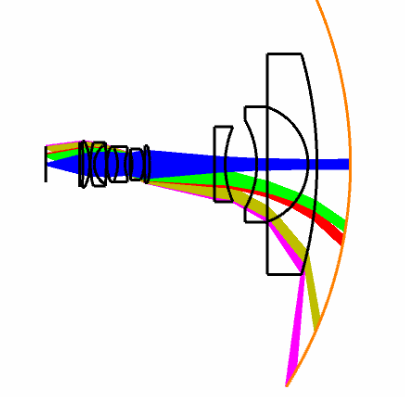}
\end{subfigure}
\caption{\label{fig:lenses} Illustration of the Ray Transfer Function (RTF). (a) The Double Gauss lens can be accurately simulated with rays represented on an input and output plane. (b) The wide angle lens can be accurately simulated with input rays on a plane and output rays on a spherical surface.  Because of the high degree of bending near the aperture edge, certain rays would not intersect any output plane to the right of the final vertex.}
 \end{figure}

Extending the usual formulation to include the z components of the origin and direction is helpful for two specific tasks.
\emph{First}, using the ray-transfer matrix, the input and output planes are fixed. Generally, we expect that the output direction vector points away from the optics so that $d_z=\sqrt{1-d_x^2-d_y^2}$. This is not true in all cases: for mirrors and wide-angle lenses $d_z$ can be negative.
\emph{Second}, including the z component allows for applying the RTF to curved output \emph{surfaces} rather than only planes (Fig.~\ref{fig:rtfcurve}). This enables us to capture rays traveling backwards or parallel to the planes which would otherwise be incorrectly represented (e.g., Fig.~\ref{fig:wideangle}).

\section{RTF for rotationally symmetric lenses with vignetting}
\begin{figure}[htpb!]
  \includegraphics[width=0.99\linewidth]{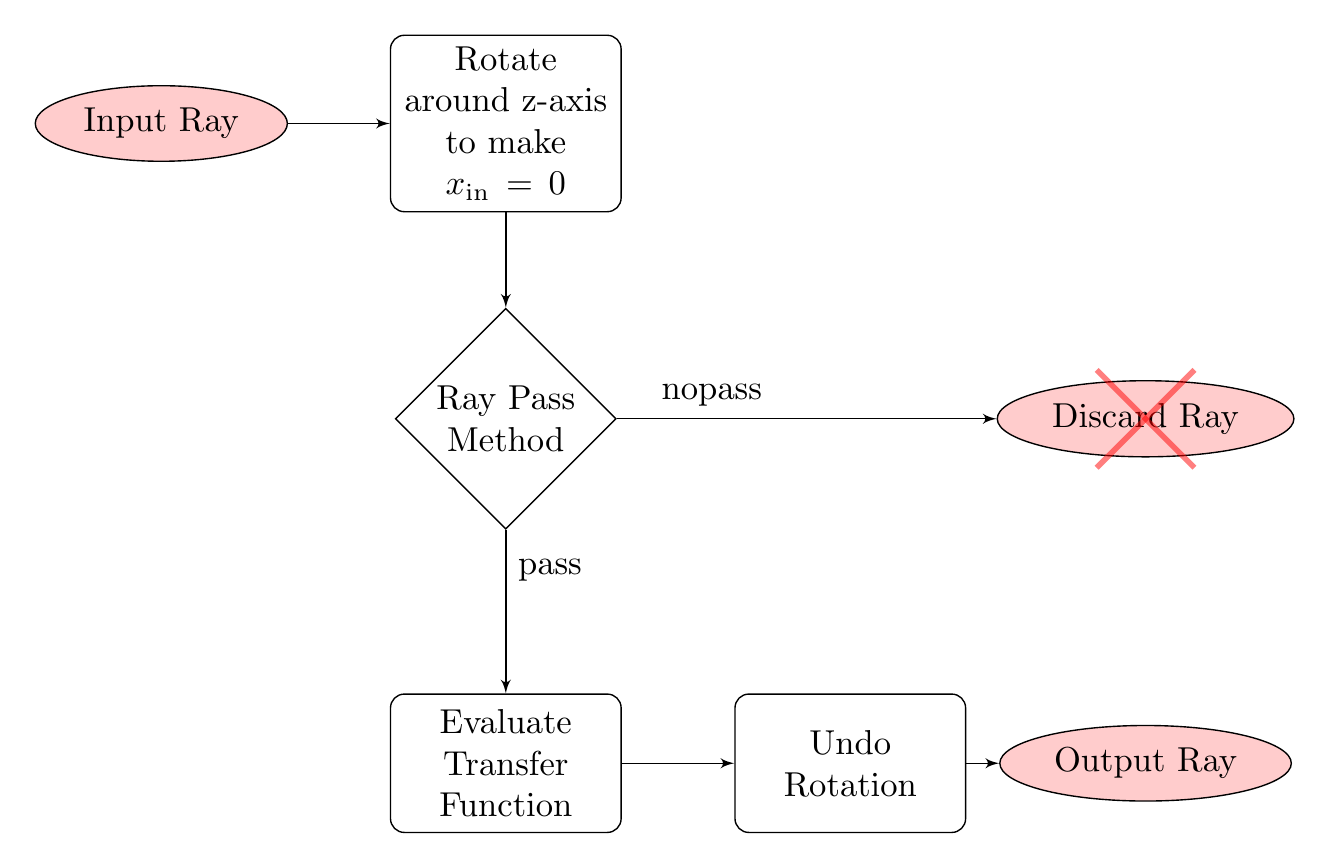}
  \caption{  \label{fig:architecture} Architecture of the ray transfer function. See text for details.}
\end{figure}

Given any particular aperture, many incoming rays will not exit the optics (f-number dependent vignetting). We implemented a method to determine whether an input ray will pass through the optics (e.g. similar to an entrance pupil).  This \emph{ray-pass method} effectively defines the valid input domain of the RTF.  The ray-transfer function converts a valid input ray to an output ray. This architecture is summarized as a flowchart in (Fig.~\ref{fig:architecture}).

\begin{figure}[t]
\includegraphics[width=0.99\linewidth]{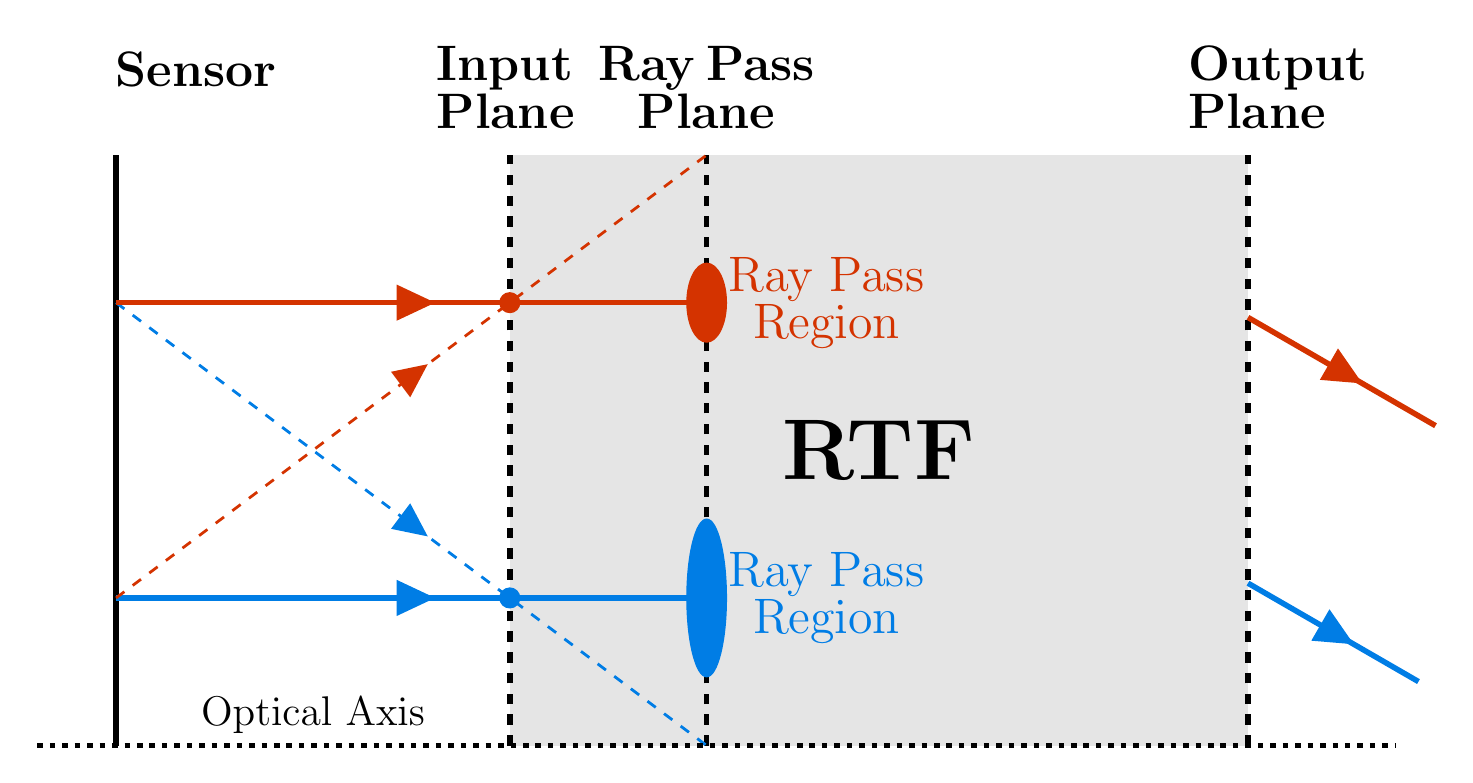}
  \caption{  \label{fig:planes} Ray-transfer function implementation in PBRT. A ray is traced from the sensor to the input plane. The position on the input plane - not the sensor plane - determines which ellipse is used.
  The ray pass method determines which rays pass, just like an entrance pupil. The dashed arrows represent rays that do not make it through the lens. }
\end{figure}

The RTF architecture for rotationally symmetric lenses consists of three auxiliary planes: the input plane, the output plane, and the ray pass plane (Fig.~\ref{fig:planes}). Rays are initially traced from sample positions in the \emph{sensor plane} to positions on the \emph{input plane}. 
The input plane is placed at a small distance from the first surface (sensor-side) of the lens, and the output plane is placed at a small distance beyond the last surface of the lens model (scene-side). Finally, the ray pass plane is placed at a nonzero distance from the input plane. A natural choice is to place the ray pass plane in the paraxial exit pupil plane. The distances between these three planes are stored in the RTF JSON file that is read by PBRT.

A PBRT simulation file contains references to all materials, lights and objects. The file also specifies the camera by defining a lens file and its position. The position of the sensor is determined by the 'film distance' parameter --- this is the distance from the sensor to the first surface of the lens:

\newpage
\begin{verbatim}
Camera "realistic" 
  "float filmdistance" [0.057315] 
  "string lensfile" "dgauss28deg.dat" 
\end{verbatim}
Likewise, we encapsulate the implementation details of the RTF in a JSON file
so that we can treat the RTF just like a normal lens in PBRT:
\begin{verbatim}
Camera "raytransfer" 
  "float filmdistance" [0.057315] 
  "string lensfile" "dgauss28deg-raytransfer.json" 
\end{verbatim}
For the RTF, the film distance measures the distance from the sensor to the first surface of the Zemax black box.
The distance can be modified without changing the ray pass method or the ray transfer function.


\section{Estimation of Ray-Transfer Functions}

There are at least two approaches to obtain an approximate ray-transfer function: expansion methods and fitting methods.
The \emph{expansion} method uses full knowledge of the lens and calculates a ray-transfer polynomial by a recursive Taylor-expansion of the ray-trace equations \cite{Hullin2012b}. The \emph{fitting} methods begins with a ray-trace dataset and finds a polynomial \cite{Zheng2017a,schrade2016} or neural network \cite{Zheng2017} as an interpolating function. Also for fitting methods, knowledge of the lens design can be incorporated. For example, in \cite{schrade2016} the final surface of a wide angle lens is used as the output surface of the polynomial.
In our work, we aim to be fully agnostic about the lens design; we fit polynomials using only the ray tracing data obtained with Zemax.


The generation of the ray dataset is a crucial step to obtain accurate results.
In what follows, we explain how we generate a dataset, exploit rotational symmetry, fit a polynomial and account for vignetting with a ``ray-pass function''. 

\subsection{Lens reversal}
For efficiency, modern ray tracing software typically follows rays from the sensor (film) plane into object space. 
Hence when obtaining the black box Zemax lens, the first step is to reverse the lens using the procedure documented in \cite{zemaxreverse}.

\subsection{Dataset generation using Zemax}
\label{sec:orga41a1be}

We generated a dataset for six lenses from the Zemax sample library: Double Gauss 28 deg, Inverse telephoto, Petzval, Cooke 40 deg, Tessar, and a Wide angle lens 200 deg. For most lenses, we placed an input and output plane at a distance of 0.01 mm from the first vertex on either side. For the wide angle 200 deg lens, we used a spherical output surface (Fig.~\ref{fig:wideangle}). This is necessary to capture all the outgoing rays. 
We experimented with different input planes positions. For example, for the Petzval lens we found that placing the planes at a 5 mm offset facilitated creating the ray pass method which varied more smoothly and thus were more accurately interpolated given a fixed sampling density. For the wide angle lens, we placed the input plane at the scene-side focal plane.

To ensure that the full exit pupil is sampled at each chosen field height, we use the pupils that Zemax calculates by default. We implemented a Zemax macro that samples the pupil using normalized pupil coordinates (Zemax terminology). For the wide angle lens, ray aiming in Zemax was enabled to more accurately estimate the pupil dimensions. The Zemax macro is included in the ISETRTF repository \cite{rtfpaper}.

For rotationally symmetric lenses, we only have to sample radially along one axis (the y-axis in our case): any incoming ray can be rotated around the optical axis (Z) such that its x-coordinate becomes zero. The output of the Zemax macro is a text file containing the correspondences between input and output rays:
\begin{table}[H]
  
  \begin{tabular}{|lll lll|lll lll| }
    \hline
    \multicolumn{6}{|c|}{Input ray} &  \multicolumn{6}{c|}{Output ray}\\
  
    $\textbf{x}$ & $\textbf{y}$ & $\textbf{z}$ & $\textbf{d}_x$ & $\textbf{d}_y$ & $\textbf{d}_z$  & $\textbf{x}$ & $\textbf{y}$ & $\textbf{z}$ & $\textbf{d}_x$ & $\textbf{d}_y$ &  $\textbf{d}_z$ \\
    \hline
     0 & $y$ & $z$ & $d_x$ & $d_y$ & $d_z$  & $x$ & $y$ & $z$ & $d_x$ & $d_y$ & $d_z$  \\
    \multicolumn{6}{|c|}{\vdots} &  \multicolumn{6}{c|}{\vdots}\\
     0 & $y$ & $z$ & $d_x$ & $d_y$ & $d_z$  & \tiny NaN & \tiny NaN & \tiny NaN & \tiny NaN &  \tiny NaN &  \tiny NaN  \\

\hline    
  \end{tabular}
 
  \caption{\label{tbl:dataset} Organization of an RTF dataset generated by the Zemax macro. Each row represents an input-output ray pair. The $(x,y,z)$ values show the position in the input and output planes. The NaNs in the output indicate that the input ray fails to exit. The input ray is only represented for $x=0$ because input rays are rotated to the y-axis for analysis (see Fig.~\ref{fig:architecture}).}
\end{table}

\subsection{Fitting the polynomial transfer function}
In its most general form, the RTF consists of six multivariate polynomials, each a function of six variables ($\mathbb{R}^6 \rightarrow \mathbb{R}^6$).
However, the input space can be reduced to $\mathbb{R}^3$ by assuming $d_Z = \sqrt{1-d_x^2-d_y^2}$  and by exploiting rotational symmetry (see also Fig.~\ref{fig:architecture}).
Including these assumptions, the transfer function calculation becomes (Fig.~\ref{fig:architecture}):
\begin{equation}
  \label{eq:rotationalsymmetry}
    \left[
    \begin{array}{l}
    \\ \bvec{p}_{\text{in}} \\\\
    \hline
    \\ \bvec{d}_{\text{in}} \\\\
  \end{array}\right]
\xrightarrow[\text{Rotation}]{}
    \left[\begin{array}{l}
    \hat{y}  \\ \hline \hat{d_{x}} \\ \hat{d_{y}}
         \end{array}\right]_{\text{in}}
\xrightarrow[\mathbb{R}^3\rightarrow\mathbb{R}^6]{\text{RTF}}
  \left[
  \begin{array}{l}
    \\ \hat{\bvec{p}}_{\text{out}} \\\\
    \hline
    \\ \hat{\bvec{d}}_{\text{out}} \\\\
  \end{array}\right]
\xrightarrow[\text{Rotation}]{\text{Undo}}
  \left[
  \begin{array}{l}
    \\ \bvec{p}_{\text{out}} \\\\
    \hline
    \\ \bvec{d}_{\text{out}} \\\\
  \end{array}\right].
\end{equation}
Not all output variables are always needed. When using output planes instead of surfaces, one can ignore the redundant output variables $z$ and $d_z$.
For curved output surfaces, we used all six output variables (Fig.~\ref{fig:wideangle}). 
More efficient ways to encode the information exist but are not explored here.

For fitting the RTF, we used a multivariate polynomial fitting toolbox from the MATLAB Exchange \cite{polyfitn}.
In Section \ref{sec:validation}, we define metrics to quantify the quality of the polynomial fit from an imaging point of view.
As we obtained satisfactory results, we did not explore other polynomial fitting algorithms as proposed in prior art \cite{Zheng2017a,schrade2016}. Overfitting is not a concern because there is no noise. 

\subsection{Accounting for vignetting: The Ray-Pass Function}
The fitted polynomial can be evaluated for all input ray positions and directions. Although the polynomial returns a value, some of the input rays are blocked (vignetted), depending on angle and field height.
Hence, we developed a ``ray-pass function'' that decides which rays should be traced (Fig.~\ref{fig:architecture}). With this architecture, only the ray-pass function must be updated when changing the f-number (diaphragm). 

\begin{figure}[t] 
  \centering 
\includegraphics[width=0.99\linewidth]{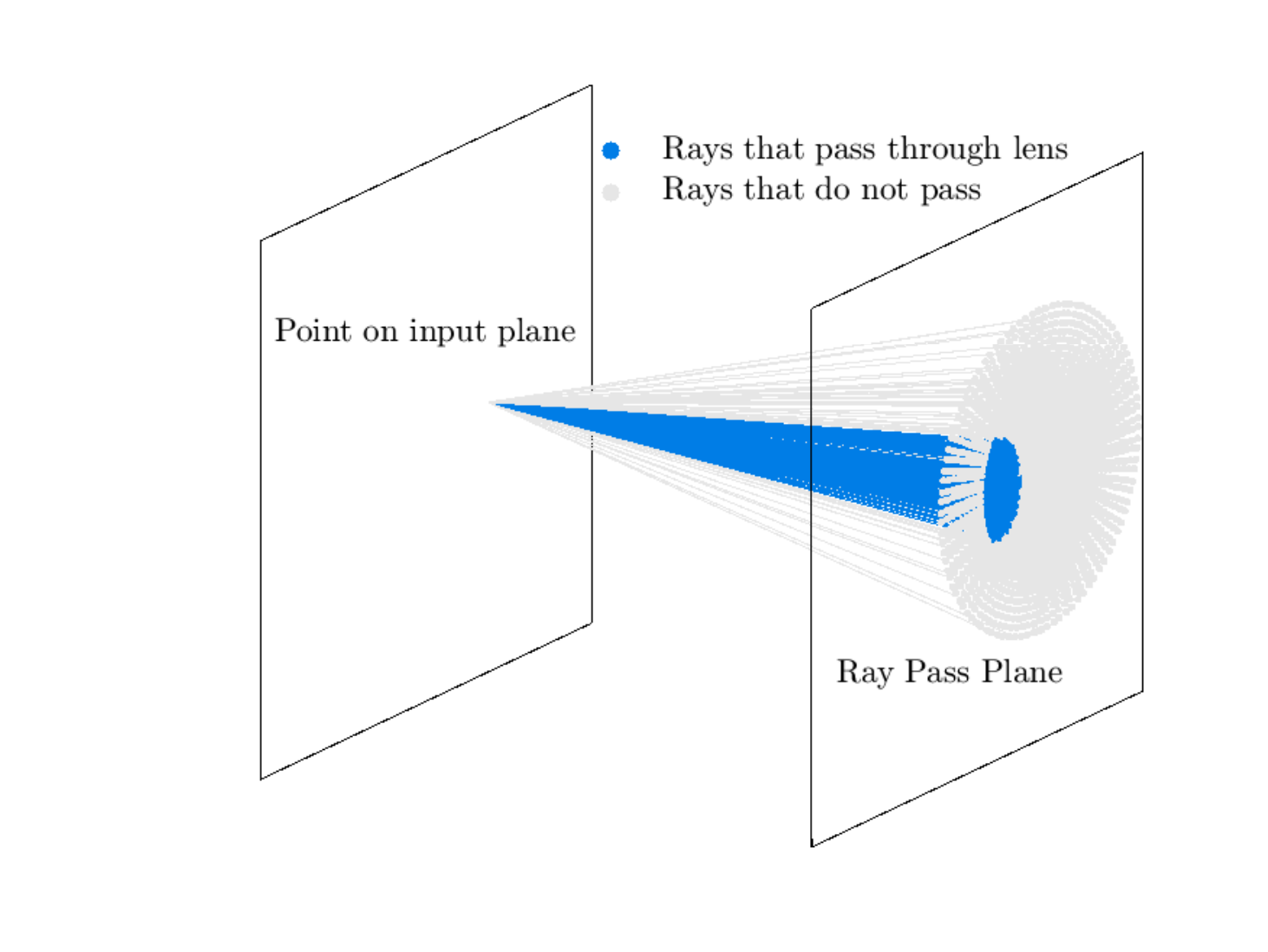}
    \caption{\label{fig:subset}Considering all the rays emerging from a position on the RTF input plane, only a subset (blue) exit from the lens.  Usually, but not always, the rays that pass through the lens define a compact region.}
\end{figure}

We implemented a method to represent the subset of passing rays as a function of field height as follows.
First, group rays originating at the same position on the input plane.
Second, calculate their intersection with an arbitrarily chosen  ray pass plane.
These intersections produce convex shapes which can be mathematically approximated as a function of field height (Fig.~\ref{fig:subset}). When the position of the ray pass plane corresponds to the exit pupil plane, the convex shapes correspond to the (effective) exit pupil as a function of field height. 
By projecting the rays onto the same plane for all positions, we do not need to explicitly model how the exit pupil can become distorted, change in orientation or position (also called pupil walking). Hence the addition of a fourth "ray pass plane" in Fig.~\ref{fig:planes}.

\begin{figure}[t]
  \centering 
\includegraphics[width=0.8\linewidth]{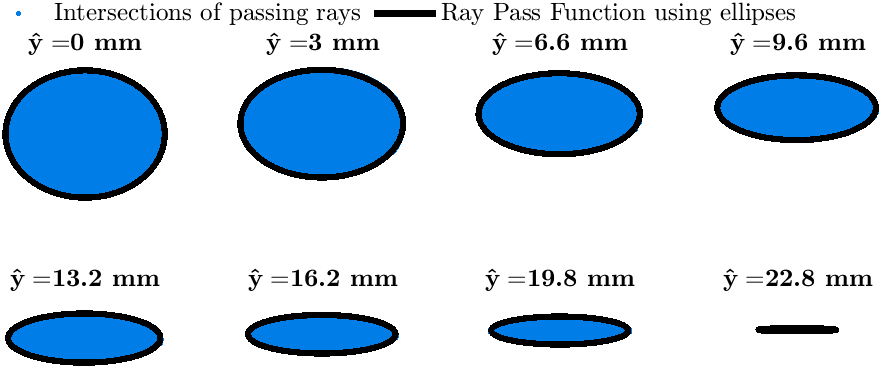}
    \caption{Ellipses are used to approximate the convex shape of the ray-pass function for rays originating at different positions $\hat{y}$ on the input plane.}
\end{figure}


Multiple choices exist to mathematically represent the pupil shapes: Circle intersections (Appendix~\ref{app:circles}), ellipses \cite{King1968}, closest neighbor masks \cite{Zheng2017a}.
In this work we implemented both the circle intersection and ellipse approach.
\emph{Circle intersections} are physically meaningful and accurate in the paraxial regime (Appendix~\ref{app:circles}). A useful feature is that changing the f-number simply corresponds to changing the size of the circle corresponding to the exit pupil. 

The main method we advocate in this work is to implement the ray pass methods using ellipses \cite{King1968}, an approach also used by Zemax (i.e., vignetting factors). While this has no physical basis, this approach is often a good approximation when the exit pupil is convex. In addition, the ellipses can robustly fit the data \cite{minvolellipse} and their parameters can be interpolated. In our software, we define vectors of ellipse radii and ellipse centers as a function input plane position. Evaluation at intermediate positions is accomplished by linear interpolation to estimate the parameters between the two nearest neighbors.



\section{Evaluation of Ray-Transfer Functions}
\label{sec:validation}

The obtained ray-transfer functions are fitted approximations, not exact representations.
We evaluate how well the approximation performs from a camera simulation perspective in several different ways.

\begin{figure}[t]
  \centering
  \begin{subfigure}[t]{0.48\linewidth}
  \vspace{0.5cm}
    \captionsetup{justification=centering}
    \caption{}
    \includegraphics[width=0.99\linewidth]{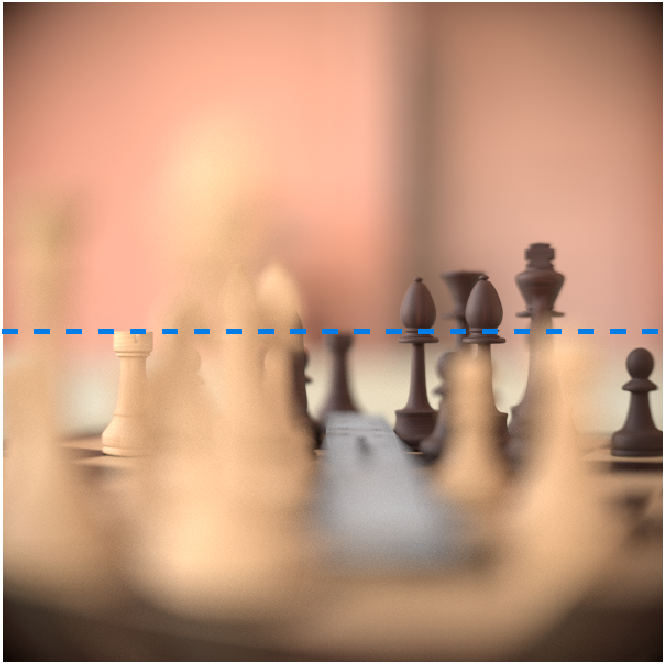}
  \end{subfigure}
  \hfill
  \begin{subfigure}[t]{0.48\linewidth}
    \captionsetup{justification=centering}
   \vspace{0.5cm}
    \caption{}
    \includegraphics[width=0.99\linewidth]{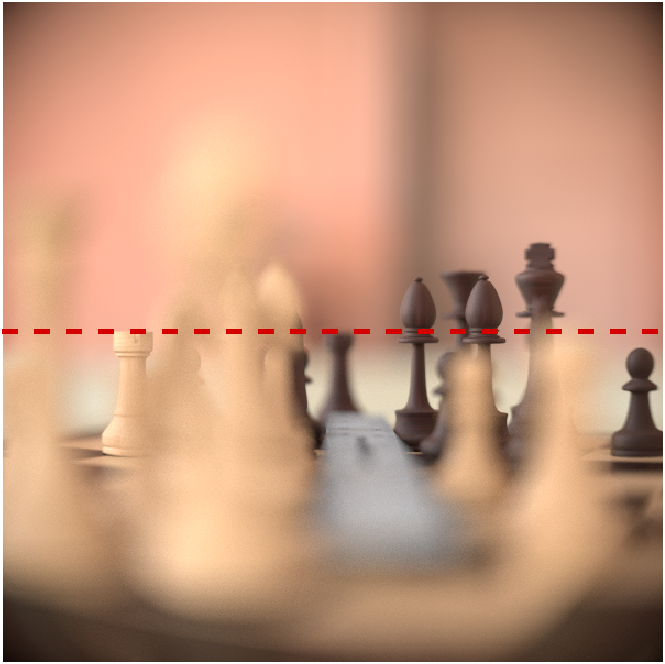}
  \end{subfigure}
  
  \caption{\label{fig:chess} A comparison of ray traced images using (a) a known lens, and (b) the RTF approximation to the lens. The images are very close to identical. Both reproduce the out-of-focus behavior accurately. The lens surfaces and refractive indices of the Double Gauss 28 deg lens are specified in the Zemax lens library. The dashed lines indicate the horizontal line analyzed in Figure~\ref{fig:linecomparisons}.}
\end{figure}

\begin{figure}[tp]
    \begin{subfigure}[t]{0.99\linewidth}
    \caption{Double Gauss 28 deg}
    \includegraphics[width=0.99\linewidth]{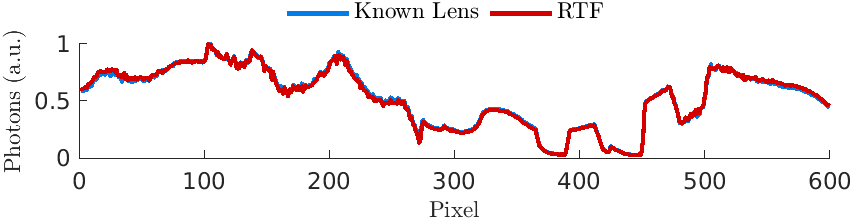}
  \end{subfigure}
      \begin{subfigure}[t]{0.99\linewidth}
   \vspace{0.5cm}
    \caption{Inverse Telephoto Lens}
    \includegraphics[width=0.99\linewidth]{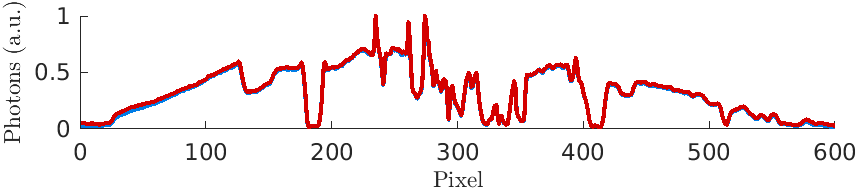}
  \end{subfigure}
      \begin{subfigure}[t]{0.99\linewidth}
   \vspace{0.5cm}
    \caption{Petzval Lens}
    \includegraphics[width=0.99\linewidth]{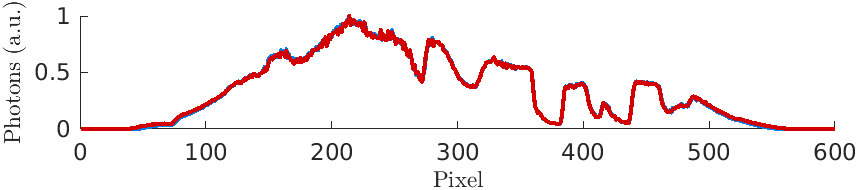}
  \end{subfigure}
        \begin{subfigure}[t]{0.99\linewidth}
   \vspace{0.5cm}
    \caption{Cooke 40 deg}
    \includegraphics[width=0.99\linewidth]{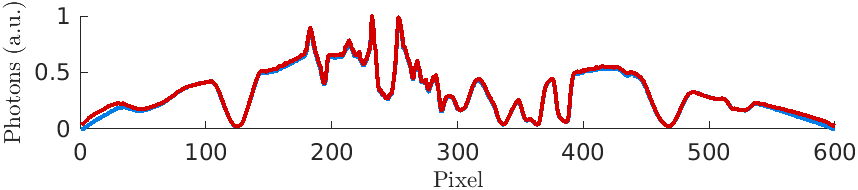}
  \end{subfigure}
        \begin{subfigure}[t]{0.99\linewidth}
   \vspace{0.5cm}
    \caption{Tessar }
    \includegraphics[width=0.99\linewidth]{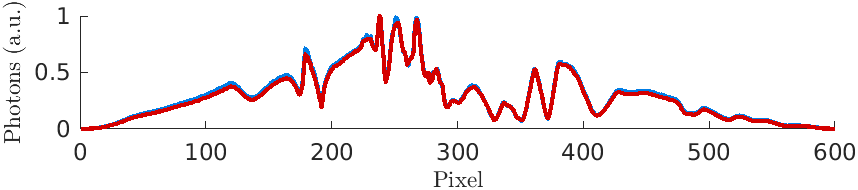}
  \end{subfigure}
        \begin{subfigure}[t]{0.99\linewidth}
   \vspace{0.5cm}
    \caption{Wide angle 200 deg}
    \includegraphics[width=0.99\linewidth]{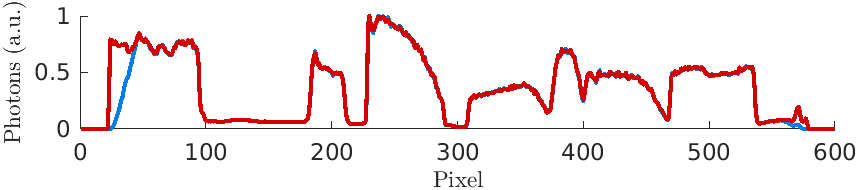}
  \end{subfigure}
  \caption{\label{fig:linecomparisons} The lines plot normalized photon count across a horizontal for rendered chess set scene (Fig.~\ref{fig:chess}). The red and blue curves compare the known (blue) and RTF (red) calculations. The wide angle lens (f) is the only case for which the image circle is not correctly predicted. We explain the reason in Appendix~\ref{app:wideangle}.}
\end{figure}

A visual comparison of a three-dimensional rendered image is a first test of the methodology but cannot be used when the lens design is actually unknown.
We implemented six lens designs from the Zemax sample library as a "realistic" lens in PBRT \cite{Kolb1995a}.
We then approximated the lenses using the RTF method and compared the rendered results for both the real lens and the RTF approximation.  The rendered images are nearly indistinguishable. See Fig.~\ref{fig:chess} for a rendered scene for the Double Gauss lens. For all other lenses we show a horizontal line through the center of the scene (Fig.~\ref{fig:linecomparisons}).
Ignoring dispersion, color was obtained using the spectrum of the light source and the reflectance of the materials.
This visual evaluation is currently not possible using only Zemax which uses a point-spread convolution approach that applies only to two-dimensional scenes.
 
To quantify the accuracy of the PBRT implementation of the RTF, we choose two metrics: the relative illumination and the edge-spread function which are compared to the prediction by Zemax.

\subsection{Relative Illumination}
The relative illumination is the radial fall-off in image irradiance; it is a common metric used for image systems evaluation. Its value is measured by imaging a uniformly illuminated Lambertian surface. The relative illumination mostly depends on the area of the exit pupil at different field heights. Hence, a good fit with Zemax indicates that the ray-pass function accurately models the vignetting.

\begin{figure}[htpb!]
\centering
  \begin{subfigure}[t]{0.32\linewidth}
          \captionsetup{justification=centering}
          \centering
        \caption{\label{fig:d}\textbf{Dbl. Gauss 28deg}}
\includegraphics[width=0.99\textwidth]{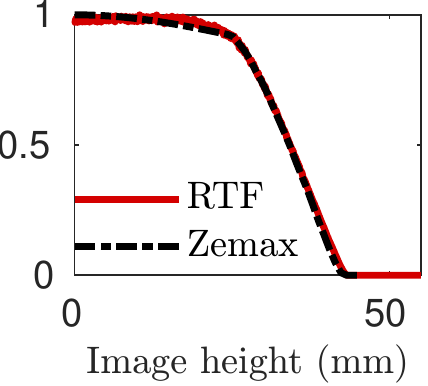}
\end{subfigure}
  \begin{subfigure}[t]{0.32\linewidth}
          \captionsetup{justification=centering}
          \centering
        \caption{\label{fig:d}\textbf{Inv. Telephoto}}
\includegraphics[width=0.99\textwidth]{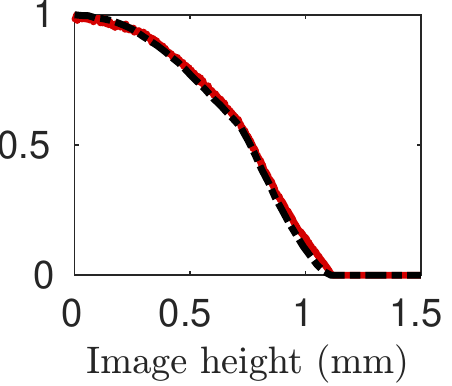}
\end{subfigure}
\begin{subfigure}[t]{0.32\linewidth}
          \captionsetup{justification=centering}
          \centering
        \caption{\label{fig:d}\textbf{Petzval}}
\includegraphics[width=0.99\textwidth]{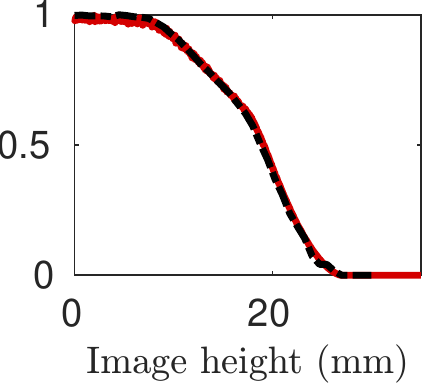}
\end{subfigure}
\begin{subfigure}[t]{0.32\linewidth}
\vspace{0.5cm}  
          \captionsetup{justification=centering}
          \centering
        \caption{\label{fig:d}\textbf{Cooke 40deg}}
\includegraphics[width=0.99\textwidth]{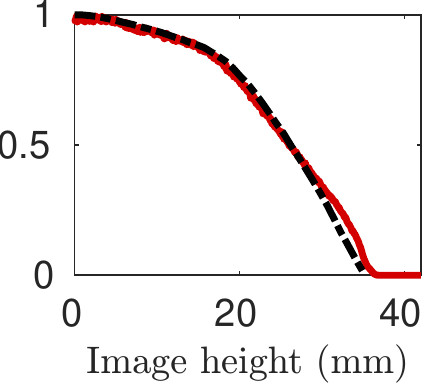}
\end{subfigure}
\begin{subfigure}[t]{0.32\linewidth}
\vspace{0.5cm}  
          \captionsetup{justification=centering}
          \centering
        \caption{\label{fig:d}\textbf{Tessar}}
\includegraphics[width=0.99\textwidth]{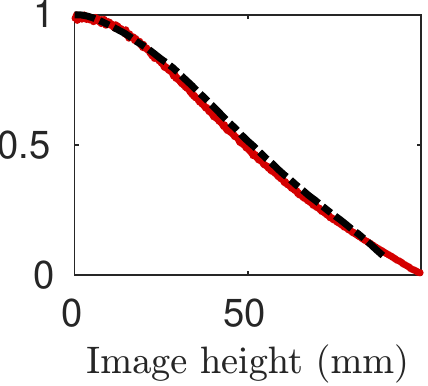}
\end{subfigure}
\begin{subfigure}[t]{0.32\linewidth}
\vspace{0.5cm}    
          \captionsetup{justification=centering}
          \centering
        \caption{\label{fig:d}\textbf{Wide-angle Lens}}
\includegraphics[width=0.99\textwidth]{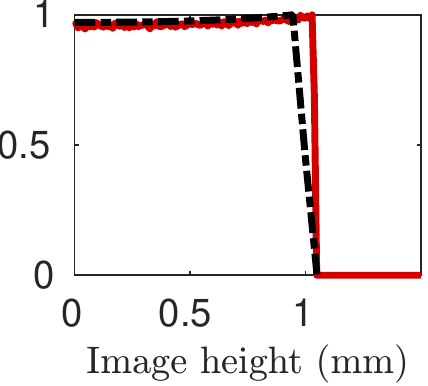}
\end{subfigure}
 \caption{\label{fig:relativeillum}Relative illumination prediction by the RTF in PBRT (red solid line) match the Zemax prediction (black dashed line). The Ray-Pass function was modeled using a sequence of field-height dependent ellipses, except for the Petzval lens which used the circle intersection method (see Appendix~\ref{app:circles}). The mismatch for the wide-angle lens is discussed in Appendix~\ref{app:wideangle}.}
\end{figure}

We calculate the relative illumination for six lenses in the Zemax sample library. In all cases there is a good match with the Zemax calculation (Fig.~\ref{fig:relativeillum}).
For the Petzval lens we found that the ellipse approximation is less suitable and that a ray pass method based on circle intersections performs much better (see Appendix~\ref{app:circles}).

\subsection{Edge Spread Functions}

To evaluate the fitted RTF from an imaging perspective we calculate its geometrical edge-spread function (ESF) and compare it to the prediction by Zemax. The ESF is the image of a step function and measures how much blurring occurs. Zemax has a convenient function to generate the geometrical ESF without diffraction so that they can be compared to the PBRT simulations, which also do not include diffraction.
To efficiently obtain the ESF in PBRT with micron resolution we render the scene for only one row of pixels on the film/sensor.

\begin{figure}[t]
  \centering
  \begin{subfigure}[t]{0.32\linewidth}
          \captionsetup{justification=centering}
          \centering
        \caption{\label{fig:d}\textbf{Dbl. Gauss 28deg}}
\includegraphics[width=0.99\textwidth]{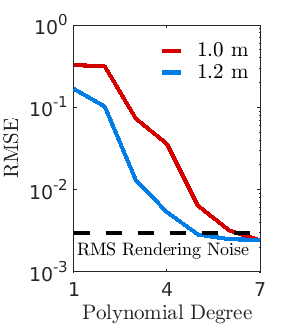}
\end{subfigure}
  \begin{subfigure}[t]{0.32\linewidth}
          \captionsetup{justification=centering}
          \centering
        \caption{\label{fig:d}\textbf{Inv. Telephoto}}
\includegraphics[width=0.99\textwidth]{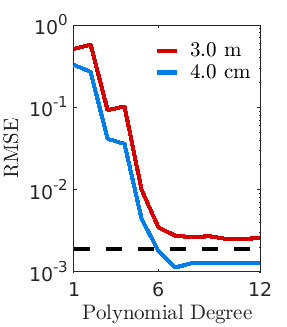}
\end{subfigure}
\begin{subfigure}[t]{0.32\linewidth}
          \captionsetup{justification=centering}
          \centering
        \caption{\label{fig:d}\textbf{Petzval}} 
\includegraphics[width=0.99\textwidth]{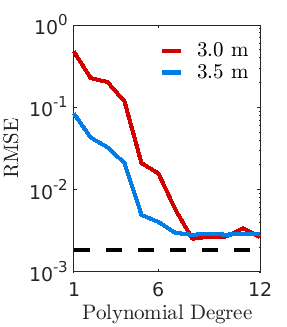}
\end{subfigure} 
\begin{subfigure}[t]{0.32\linewidth}
\vspace{0.5cm}  
          \captionsetup{justification=centering}
          \centering
        \caption{\label{fig:d}\textbf{Cooke 40deg}}
\includegraphics[width=0.99\textwidth]{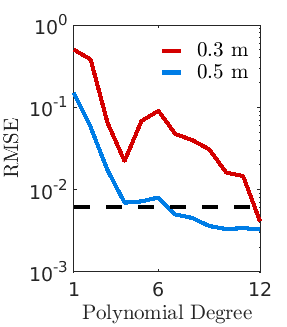}
\end{subfigure}
\begin{subfigure}[t]{0.32\linewidth}
\vspace{0.5cm}  
          \captionsetup{justification=centering}
          \centering
        \caption{\label{fig:d}\textbf{Tessar}} 
\includegraphics[width=0.99\textwidth]{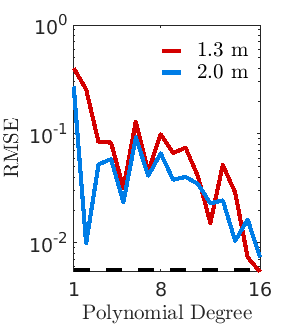}
\end{subfigure}
\begin{subfigure}[t]{0.32\linewidth}
\vspace{0.5cm}    
          \captionsetup{justification=centering}
          \centering
        \caption{\label{fig:d}\textbf{Wide-angle Lens}} 
\includegraphics[width=0.99\textwidth]{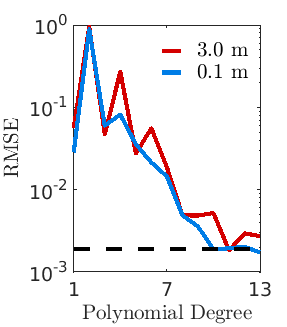} 
\end{subfigure}
\caption{\label{fig:esf-rms} The difference ($\log_{10}$ of the root mean squared error, RMSE) between the RTF and Zemax edge-spread functions as a function of polynomial degree. The two curves in each panel show different object distances (measured from front vertex). As polynomial degree increases, RMSE generally declines. The horizontal dashed lines represent the rendering noise from ray tracing (see text for details). For all lenses, the RMSE approaches the level of the rendering noise.}
\end{figure}

To select the polynomial degree for the RTF, we compared the root mean squared error (RMSE) between RTF approximation and the Zemax edge spread functions.  The smallest achievable level is defined by the rendering noise of the PBRT simulation. We calculate this noise by rendering a perfectly uniform input scene with the same rendering parameters (number of rays, ray bounces, path tracer, and lens). The standard deviation (RMSE) of the rendered uniform scene is used as an estimate of the rendering noise.  The RMSE generally declines with increasing polynomial degree, becoming very close to the PBRT rendering noise (Fig.~\ref{fig:esf-rms}).

\begin{figure}[htpb!]
\centering
  \begin{subfigure}[t]{0.32\linewidth}
          \captionsetup{justification=centering} 
          \centering
        \caption{\label{fig:d}\textbf{Dbl. Gauss 28deg}}
\includegraphics[width=0.99\textwidth]{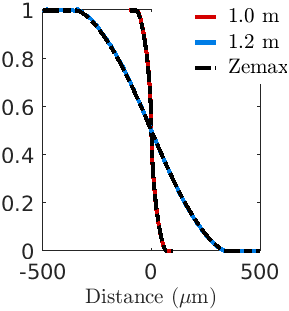}
\end{subfigure}
  \begin{subfigure}[t]{0.32\linewidth} 
          \captionsetup{justification=centering}
          \centering
        \caption{\label{fig:d}\textbf{Inv. Telephoto}}
\includegraphics[width=0.99\textwidth]{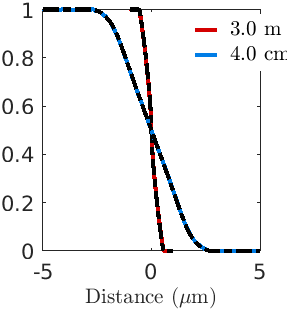}
\end{subfigure}
\begin{subfigure}[t]{0.32\linewidth}
          \captionsetup{justification=centering}
          \centering 
        \caption{\label{fig:d}\textbf{Petzval}} 
\includegraphics[width=0.99\textwidth]{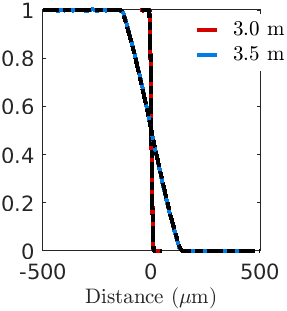}
\end{subfigure} 
\begin{subfigure}[t]{0.32\linewidth}
\vspace{0.5cm}  
          \captionsetup{justification=centering}
          \centering
        \caption{\label{fig:d}\textbf{Cooke 40deg}}
\includegraphics[width=0.99\textwidth]{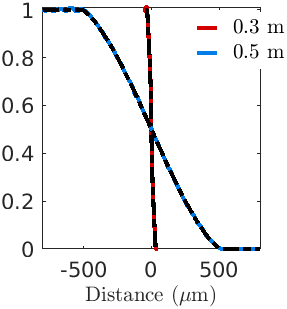}
\end{subfigure}
\begin{subfigure}[t]{0.32\linewidth}
\vspace{0.5cm}  
          \captionsetup{justification=centering}
          \centering
        \caption{\label{fig:d}\textbf{Tessar}} 
\includegraphics[width=0.99\textwidth]{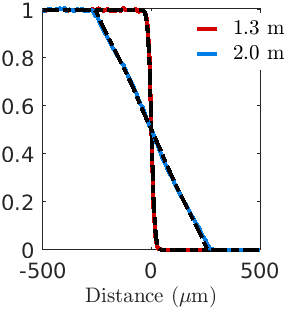}
\end{subfigure}
\begin{subfigure}[t]{0.32\linewidth}
\vspace{0.5cm}    
          \captionsetup{justification=centering}
          \centering
        \caption{\label{fig:d}\textbf{Wide-angle Lens}} 
\includegraphics[width=0.99\textwidth]{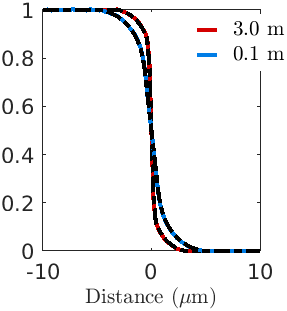} 
\end{subfigure}
\caption{\label{fig:esf}Comparing the edge-spread functions computed using Zemax and the RTF simulations implemented in PBRT. The comparison is calculated for two object distances (see panel legends). There is no substantial difference between the two methods.  The polynomial degree used to fit the RTFs differed between the lenses, and the rationale for selecting the values is shown in Fig.~\ref{fig:esf-rms}.}
\end{figure}

After selecting the polynomial degrees and ray-pass functions for each lens, we computed the edge-spread functions for the six lenses and two object distances. We compare the RTF and Zemax edge-spread functions in Fig.~\ref{fig:esf}. The agreement between the two methods is very good. This confirms that the small RMSE differences are not significant for sensor prototyping or machine learning applications.

\section{Discussion}

\subsection{Related work}
The ray transfer function is a fundamental concept in optics. The linear paraxial approximation (ABCD) is well-known from Gauss' work \cite{Gauss_1845-ds}. Extensions to the paraxial approximations of light wavefronts, rather than rays, were developed by Seidel \cite{Seidel1857-vr}. Polynomial characterizations of wavefront aberrations on circular support were introduced by Zernike \cite{Zernike1934-gn}.  These concepts are part of the classical optics toolkit. 

In certain cases, the dominant rendering properties of camera optics can be captured by the ray representation without reference to wavefronts.  Tools that extend the linear ABCD model for rays to nonlinear polynomial ray transfer approximations were developed to computer graphics and ray tracing by Hullin et al. \cite{hullin2012polynomial}, who provided software to create polynomial ray transfer functions. They also described applications such as modeling lens flare. Several papers built on this work, describing efficient ways to find and implement the polynomial approximations and considering alternative formulations to expand the range of lenses that could be accurately modeled \cite{schrade2016sparse,Zheng2017,Zheng2017a}. The metrics used to evaluate the polynomial models included examples of rendered images and metrics in the ray parameter space. 

This paper builds on the existing polynomial framework in several ways. A main motivation is to implement methods that can be applied even when the specific lens design is not known. Construction of the polynomials from the black box description extends the domain of application and is particularly important for soft prototyping applications (e.g., \cite{lyu2022arxiv}). Second, we evaluate the accuracy of the polynomial models using image domain metrics: the relative illumination and edge-spread functions. Evaluating the polynomial approximation in the ray representation is valuable, but in many applications the critical question is whether the soft prototype of machine learning algorithm has the correct sensor data.  Hence, evaluation of the accuracy with respect to the spectral irradiance and thus the sensor response is important. Third, we provide open-source software to enable others to check our work and integrate the methods into two commonly used tools: lens design software (Zemax) and physically based ray-tracing (PBRT).


\subsection{Additional advice to avoid headaches}
When generating the dataset using the attached Zemax macro, one has to be careful about several caveats. 
First, it is important to make sure the primary wavelength in Zemax equals the wavelength used for data collection as this affects some Zemax calculations like pupil estimation. Second, when calculating the relative illumination and edge-spread functions with Zemax, make sure to select a single wavelength. In addition, as recommended by Zemax, we had to turn off vignetting factors for the Tessar lens to get a satisfactory fit.
Third, turning on ray aiming is advised when expecting large distortions of the exit pupil, like for a wide-angle lens \cite{zemaxrayaim}. This avoids undersampling of the exit pupil.

In the case of wide-angle lenses, high degree polynomials are required to make the RTF fit well across the whole image circle.
By placing the input plane near the position of the sensor (or the focal plane) image circle of interest can be controlled by limiting the input radius. In this case, the polynomial only needs to predict the selected region which provides a better fit at lower degrees with fewer coefficients. We verified that, for example, for an image circle radius of 0.8 mm (instead of 1.04), a fifth degree polynomial can already match the edge spread functions.

\subsection{Future work}

Prior work models dispersion by using wavelength as a variable in the polynomials. Instead, we implemented dispersion by fitting an RTF for each wavelength of interest. We will validate this approach in future work.

The method illustrated here and implemented in the software applies only to rotationally symmetric lenses.
This method is satisfactory for the vast majority of lenses, but freeform lenses are becoming more common. Sampling and fitting an RTF for these lenses in full 6D space (position and direction) is a significantly larger computational problem, as is ray tracing for free form lenses. We do not have examples of such lenses, but we consider algorithm work on this topic as a something for the future.

It may be useful to distinguish between the rotational symmetry of the the polynomial transfer function and the ray-pass function. For example, the lens surfaces might be spherical but cut asymmetrically (e.g. square shape). In addition, the diaphragm, or a coded aperture might not be rotationally symmetric.
In such a case, the fitted polynomial can continue to exploit rotational symmetry while the ray-pass function could be generalized. 

Another possible extension is to include ray-based diffraction techniques.
While fundamentally ray-tracing approaches do not simulate wave-optics phenomena, useful methods for approximating diffraction with ray-based diffraction techniques have been proposed \cite{Lin2014,Freniere1999,Lian2019,Mout2016}.
It would be interesting to explore how these methods could combined with a fitted ray-transfer function.

For simulations involving  diffraction, coatings, or time-of-flight, gated-cmos depth imaging, knowing the optical path lengths of the rays and their change in spectral radiance might be required. One could potentially models these effects using two additional fitted functions: a "Pathlength Function" and a "Ray Amplitude Function''. 
 
\subsection{Performance improvements}

The current PBRT implementation is several times slower than ray-tracing through the actual lens. The bulk of the calculation involves the evaluation of high-degree polynomials.  We are confident that the calculations can be significantly accelerated.  For example, similar work on ray-transfer functions used polynomials as a faster alternative for brute force ray-tracing by limiting their search to sparse multivariate polynomials \cite{schrade2016,Zheng2017a}. Readers who are concerned about speed might implement our solution using these alternative fitting algorithms.
Also, lower degree polynomials might also be suitable for simulating cameras large pixels. In this case, the edge-spread function might not be fully dominated by the optics alone and a lower-degree polynomial fit might be satisfactory. Finally, polynomial calculations can take advantage of GPUs \cite{Hullin2012b}, and we are considering implementing this method .

\section{Conclusions}
\label{sec:org2f190d7}

The methods described here significantly extend the range of camera applications that can benefit from soft prototyping. Prototyping can advance while permitting lens manufacturers to maintain the intellectual property of the lens design.  By providing a black box model, customers can build a ray-transfer function that can be incorporated into ray tracing software. We demonstrate how to integrate these functions into PBRT for the specific case of Zemax, and we show that for six very different lenses the PBRT implementation replicates the Zemax specification. In a separate paper, we use this method as part of a general simulation pipeline, comparing the simulations with data acquired from a real camera with an unknown lens design and a constructed three-dimensional scene \cite{lyu2022arxiv}.

\section*{Acknowledgements}
We thank David Cardinal and Zhenyi Liu for their contributions to the software and discussions.  We thank Google and Ricardo Motta for support via the Stanford Center for Image Systems Engineering faculty-mentor-advisor program.  Thomas Goossens holds a fellowship from the Belgian American Educational Foundation (B.A.E.F).

\section*{Data availability}
All software developed for this paper is available on GitHub.
The implementation of the ray-transfer function can be found on \url{https://github.com/scienstanford/pbrt-v3-spectral}.
The scripts to generate the ray-transfer function based on the data from Zemax can be found on \url{https://github.com/ISET/isetrtf}.

The software is intended to be used as follows.
First, one uses the Zemax macro to generate the text file containing input and output rays. Second, this dataset is processed using a MATLAB script which generates a JSON file. Lastly, this JSON file is passed as input to the customized version of PBRT.

\bibliography{referencesThomas,rtfReferences,referencesWebsites}

\appendix

\section{Ray pass methods using circle intersections}
\label{app:circles}

\begin{figure}[H]
  \begin{subfigure}[t]{0.49\linewidth}
  \includegraphics[width=0.99\linewidth,page=1]{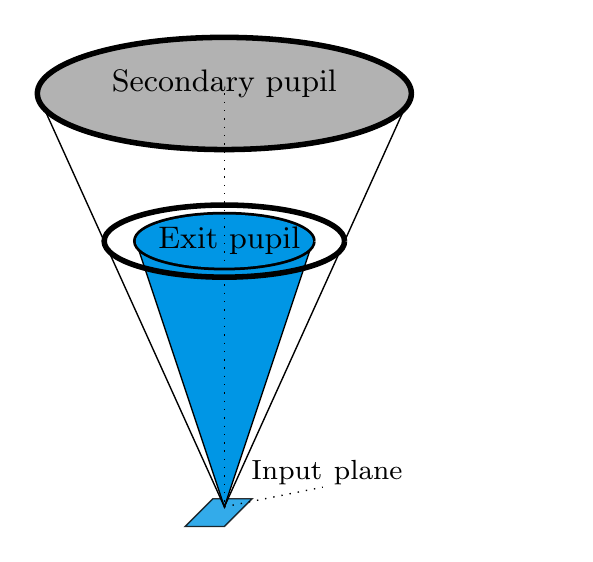}
  \end{subfigure}
    \begin{subfigure}[t]{0.49\linewidth}
      \includegraphics[width=0.99\linewidth,page=2]{./fig/tex/pdf/circleintersectionpupils}

    \end{subfigure}
 \caption{\label{fig:circles} Modeling vignetting with circle intersections by projecting the circular pupils to a common plane. Adapted from \cite[p.86]{GoossensThesis}.}
\end{figure}

Each lens and diaphragm in a complex lens can be considered a finite aperture that limits the size of the light beam. The most limiting aperture is often called the Aperture Stop.
 We can calculate the image of each aperture (called pupils) as seen from different positions on the input plane. By construction, for a ray to pass through the system, it must pass through all of the pupils.  If the diaphragms are circular, the paraxial images will be approximately circular and the shape of the lens system's pass function can be approximately modeled by the intersection of these circles \cite{Asada1996} (Fig.~\ref{fig:circles}).
 
In our work we project the diaphragms to the "ray pass plane" (Fig.~\ref{fig:planes}).
The projected circles have radii $R_i,\,i=1,2,\hdots\,$. Their displacement from the center $d_i$ depends on the input plane position $\hat{y}$ (Fig.~\ref{fig:circles}). Assuming linearity, 
\begin{equation}
    d_i = s_i \hat{y},
\end{equation}
with $s_i$ being a sensitivity parameter.

\begin{figure}[hb!]
  \includegraphics[width=0.99\linewidth]{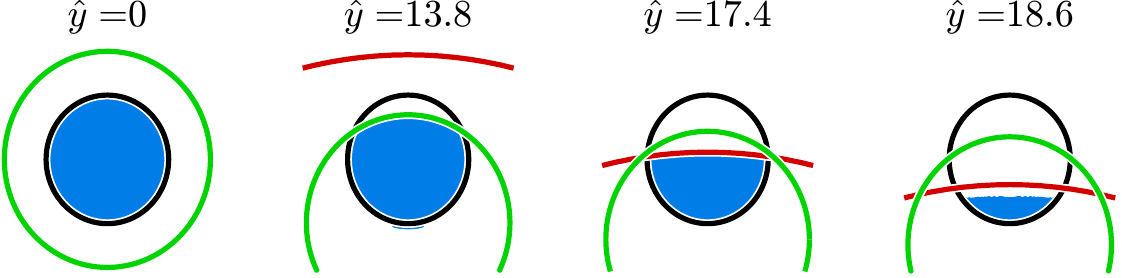}
  \caption{ The ray pass region for the Petzval lens is well modeled by intersecting three moving circles. Here shown for three position on the input plane.\label{fig:petzvalcircles}}
  \end{figure}

\begin{figure}[t]
  \centering
     \begin{subfigure}[t]{0.33\linewidth}
         \captionsetup{justification=centering}
               \caption{}
      \includegraphics[width=0.99\linewidth]{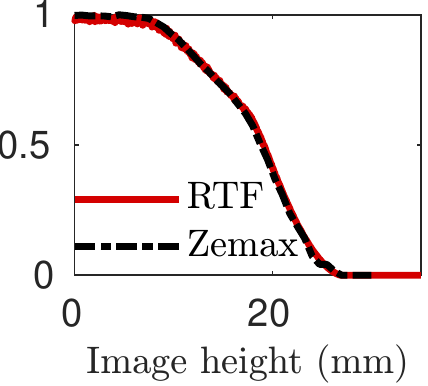}
    \end{subfigure}
    \hspace{1cm}
   \begin{subfigure}[t]{0.33\linewidth}
       \captionsetup{justification=centering}
             \caption{}
      \includegraphics[width=0.99\linewidth]{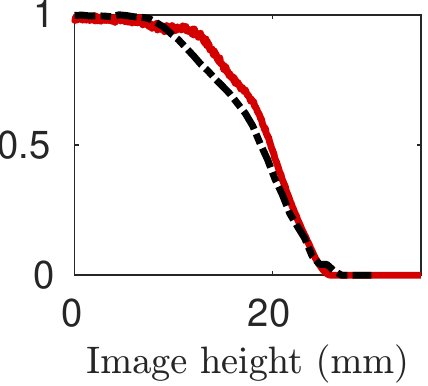}

    \end{subfigure}

    \caption{\label{fig:relativepetzval} For the Petzval lens, a ray pass method based on (a) circle intersections performs better, than (b) using ellipse approximations.}
\end{figure}

For the classic Petzval lens we require three circles to model that the ray pass region changes with input plane position (Fig.~\ref{fig:petzvalcircles}). Different diaphragms (circles) become dominant at different field heights (Fig.~\ref{fig:relativepetzval}).
Placing, arbitrarily, the ray pass plane 17 mm above the input plane, the three circles have radii 5.74 mm, 42.67 mm and 9.65 mm with corresponding sensitivities of 0.72, -1.7 and  0.30.

     
The circle radii and sensitivities are estimated in several steps. First, we select off-axis positions at which a different circle cuts the pupil (e.g., Fig.~\ref{fig:petzvalcircles}). Second, at each pupil edge we find the tangent circle with the minimal radius that encompasses all points of intersection. The sensitivity is then obtained by observing that $d_i = (\text{position on edge}-R_i)$ such that
 $s_i=d_i/\hat{y}$. This algorithm is implemented in the  \emph{findCuttingCircleEdge(..)} function in \cite{rtfpaper}.
 The script used to generate Fig.~\ref{fig:petzvalcircles} is \emph{generatePetzvalCircles.m}.

\section{Wide angle lens relative illumination}
\label{app:wideangle}
\begin{figure}[ht!]
    \begin{subfigure}[t]{0.4\linewidth}
  \includegraphics[width=0.99\linewidth]{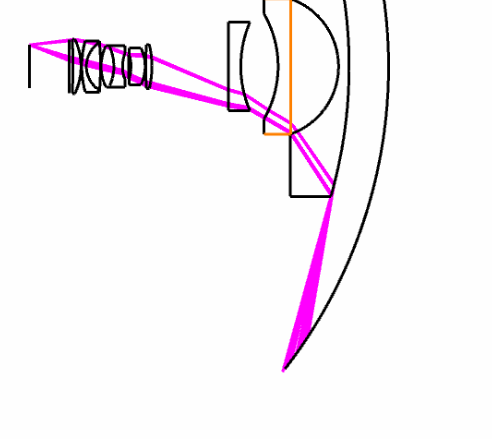}
  \end{subfigure}
    \begin{subfigure}[t]{0.49\linewidth}
  \includegraphics[width=0.99\linewidth]{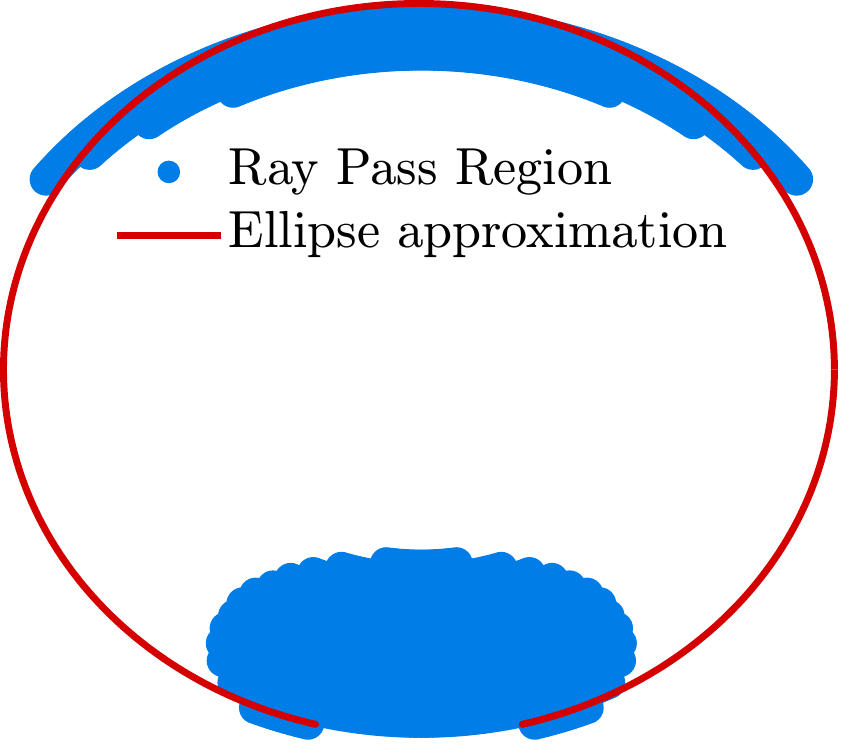}
 \end{subfigure}
  \caption{The ray pass region of the wide angle lens is not convex; characterizing the region with an ellipse overestimates the region of passing rays. Consequently, many input rays are traced that should not be included, explaining the imperfections in Fig.~\ref{fig:linecomparisons}f and \ref{fig:relativeillum}f. \label{fig:nonconvex}}
  \end{figure}

We briefly comment on the difficulty of correctly predicting the image circle radius for the wide angle lens in (Fig.~\ref{fig:linecomparisons}).
Near the edge the ray pass region becomes non-convex and consists of two  clusters (Fig.~\ref{fig:nonconvex}). An ellipse significantly overestimates the area causing the image circle to appear larger. More versatile fitting functions, like neural networks, might be suitable for this situation.

\end{document}